\documentclass[preprint1]{aastex631}

\accepted{by PSJ, May 23, 2023}

\shorttitle{Orbits \& Masses Satellites Pluto}
\shortauthors{Porter \& Canup}

\begin{document}

\title{Orbits and Masses of the Small Satellites of Pluto}

\author[0000-0003-0333-6055]{Simon B. Porter}
\email{simon.porter@swri.org}
\affiliation{Southwest Research Institute, 1050 Walnut Street, Suite 300, Boulder, Colorado, USA}

\author[0000-0002-2342-3458]{Robin M. Canup}
\affiliation{Southwest Research Institute, 1050 Walnut Street, Suite 300, Boulder, Colorado, USA}

\begin{abstract}

We present a new orbit and mass solution for the four small satellites of Pluto: Styx, Nix, Kerberos, and Hydra.
We have reanalyzed all available observations of the Pluto system obtained by the Hubble Space Telescope (HST)
from 2005 to 2019 with the ACS, WFPC2, and WFC3 instruments, 
as well as the New Horizons LORRI images taken on approach to Pluto in 2015.
We have used this high precision astrometry to produce updated orbits and mass estimates with uncertainties
for all four of the small satellites.
We find that the masses of Nix and Hydra are smaller than previously published estimates, with a dynamical
mass of 
    1.8$\pm$0.4$\times$10$^{-3}$ km$^3$/s$^2$ (2.7$\pm$0.6$\times$10$^{16}$ kg) for Nix 
and 2.0$\pm$0.2$\times$10$^{-3}$ km$^3$/s$^2$ (3.0$\pm$0.3$\times$10$^{16}$ kg) for Hydra.
These masses are 60\% and 63\% of the mean estimates by Brozovic et al. (2015), although still  consistent with their 1-sigma uncertainties, and correspond to densities of 
1.0$\pm$0.2 g/cm$^3$ for Nix and 1.2$\pm$0.2 g/cm$^3$ for Hydra given the moon volume estimates
from Porter et al (2021).
Although these densities are consistent with a range of ice-rock compositions, depending on the unknown bulk porosity in the moon interiors, the moons' high albedos and predominantly icy surfaces are most easily explained if their interiors are ice-rich.  The tiny masses of Kerberos and Sytx remain very poorly constrained; we find 1-$\sigma$ upper limits for the dynamical mass of Styx to be 3$\times$10$^{-5}$ km$^3$/s$^2$ (5$\times$10$^{14}$ kg) 
and for Kerberos 5$\times$10$^{-5}$ km$^3$/s$^2$ (8$\times$10$^{14}$ kg), 
consistent with densities of $<$2.1 g/cm$^3$ for both bodies.

\end{abstract}


\section{Background and Motivation}

Pluto and its satellites comprise one of the most interesting and complex systems of bodies in the Solar System. 
The major bodies Pluto (diameter 2374 km) and Charon (diameter 1212 km) orbit their common barycenter, 
roughly 200 km above the surface of Pluto \citep{2015Sci...350.1815S}. 
Their mutual orbit is circular to the precision that it can be measured \citep{2012AJ....144...15B} 
and both bodies are tidally locked to each other, creating a doubly-synchronous system with a period of $6.3872273\pm0.0000003$
days \citep{2012AJ....144...15B}. 
Four much smaller satellites orbit exterior to Charon: Styx with equivalent diameter $\approx$10.5 km, 
Nix with full length dimensions 48.4$\times$33.8$\times$31.4  km, 
Kerberos at 19$\times$10$\times$9 km, 
and Hydra with a size of 52.0$\times$36.5$\times$29.3 \citep{2021porterChapter}. 
These sizes correspond to best-fit approximate shapes, 
which are likely quite accurate for Nix due to nearly complete coverage of its surface by New Horizons observations, 
moderately accurate for Hydra, and only approximate for Kerberos (seen in only 2 resolved New Horizons images).
The orbits of the small satellites have low but non-zero eccentricity with respect to the system barycenter ($e<0.03$), 
are close to coplanar (but not quite coplanar) with the Pluto-Charon orbit ($I<0.5^\circ$), 
and are close to (but not quite at) mean motion commensurabilities with the Pluto-Charon orbit at 3.269:1, 3.994:1, 5.128:1, 
and 6.066:1 (see below).
Additional satellites could be stable within the orbit of Charon \citep{2013MNRAS.430.1892G,2014MNRAS.439.3300G}, but none have thus far been discovered \citep{2015Sci...350.1815S}. 
Despite extensive searches from stellar occultations \citep{2014A&A...561A.144B}, 
the Hubble Space Telescope \citep[HST;][]{2007AJ....133.1485S}, 
the Herschel Space Telescope \citep{2015A&A...579L...9M},
and the New Horizons spacecraft \citep{2018Icar..301..155L}, 
no rings have been discovered in the Pluto system, nor any additional bodies smaller than Styx.

Prior to the discovery of Nix and Hydra, \citet{2005Sci...307..546C}, building on the original impact hypothesis of
\citet{1989ApJ...344L..41M}, proposed that the Pluto-Charon system formed
as result of a low speed giant impact between two similarly sized bodies, with the impactor having 30 to 50\% of Pluto's current mass,
and showed through hydrodynamic simulations that Charon could have formed either from a disk of material around Pluto produced by the impact, or as an intact moon as a direct result of the impact, with the latter appearing more probable given Charon's mass and bulk composition. 
After the discovery of Nix and Hydra, \citet{2006Natur.439..946S} proposed that they were remnants of a
circum-Pluto disk produced by the Charon-forming giant impact.
The very low eccentricities and inclinations of the orbits of the small satellites make it most plausible that
they were formed around Pluto, but whether they were a direct product of the Charon-forming event or represent a later process or addition of material remains unclear (e.g., Canup et al. 2021 and references therein).
\citet{2011AJ....141...35C} extended the earlier hydrodynamic simulations of the Charon-forming impact to test if it was 
possible for Nix and Hydra to be circum-Pluto disk remnants while Charon was formed intact.
\citet{2011AJ....141...35C} found that Nix and Hydra should have comparable or lower mean densities than Pluto or Charon, as 
the disk was often
comprised preferentially of water ice from the mantles of the two large impactors, which would have been partially
differentiated into water ice mantles and more rock-rich cores
for scenarios with a rock-rich, massive Charon.
\citet{2010ApJ...714.1789L} showed that the Haumea system (with two satellites slightly larger than Nix and Hydra and a ring system) 
could have been formed by a similar impact to the disk-producing impacts in \citet{2005Sci...307..546C}, but with slightly different impact conditions.
The members of the collisional family associated with Haumea have strong water ice spectral features, implying that they were
water ice mantle fragments \citep{2007AJ....134.2160R}.
The spectral observations of Nix and Hydra by New Horizons were also dominated by water ice 
\citep{2018Icar..315...30C}.
However, spectral measurements only show surface composition, not bulk composition.
Understanding the density of the small satellites of Pluto can thus help to constrain not only the conditions of the 
Charon-forming giant impact, but also potentially help to constrain the state of differentiation of icy dwarf planets in the 
early Solar System.

All four of the small satellites of Pluto were discovered with imaging from HST. 
Nix and Hydra (provisionally designated ``P2'' and ``P3'', where ``P1'' was Charon) were discovered by \citet{2006Natur.439..943W} 
in images obtained by HST. 
Kerberos (provisionally ``P4'') was discovered by \citet{2011IAUC.9221....1S} 
in HST images, and then finally Styx (``P5'') by \citet{2012IAUC.9253....1S}, again in HST images. 
The small satellites of Pluto are sufficiently dim (V$\approx$23-27) and close to Pluto 
(1.8-2.8 arcseconds from the system barycenter at maximum) that space telescopes are the only practical way to observe them from the Earth. 
Since the discoveries of the small satellites, Pluto has been extensively imaged by HST, 
both before and after the New Horizons flyby. 

The other important observations of the small satellites of Pluto are from NASA's New Horizons spacecraft. 
New Horizons is a small spacecraft (478 kg at launch) that was designed to reach Pluto as fast as possible 
\citep{2008SSRv..140...49G}. 
It launched on January 19, 2006, half a year after the discovery of Nix and Hydra, whose names were selected in part as a tribute to the mission. 
After a Jupiter gravity assist flyby on February 28, 2007, New Horizons flew through the Pluto system on July 14, 2015, 
and then continued on to encounter the small Kuiper Belt Object (KBO) (486958) Arrokoth (formerly 2014 MU$_{69}$) on January 1, 2019
\citep{2019Sci...364.9771A}. 
New Horizons has two primary imaging instruments, the LOng Range Reconnaissance Imager \citep[LORRI;][]{2008SSRv..140..189C}, 
and the Multispectral Visible Imaging Camera (MVIC), part of the Ralph instrument \citep{2008SSRv..140..129R}.
LORRI is a panchromatic framing camera with a 20.8 cm aperture \citep{2008SSRv..140..189C}, 
while MVIC has 7.5 cm aperture and is primarily used as a Time Domain Integration (TDI) imager with four color filters and a 
panchromatic channel \citep{2008SSRv..140..129R}. 
While MVIC was used to image the colors of the small satellites during the flyby, and to obtain the closest image of 
Nix in panchromatic mode, it was only used to image small satellites in the few days before closest approach 
\citep{2016Sci...351.0030W}. 
LORRI began imaging the Pluto system in November 2006, shortly after launch. 
Nix and Hydra were first tentatively seen in LORRI optical navigation (\verb|OPNAV|) images in January 2015, 
about half a year before the flyby. 
The \verb|OPNAV| images continued until the flyby, and in the month leading to the flyby were joined by the ``hazards'' images, 
a series of seven deep image sequences that were designed to search for additional satellites or rings that would have been a 
hazard to the spacecraft \citep{2018Icar..301..155L}. 
While the HST images cover a much longer timebase (2005-2019), the roughly daily cadence of the LORRI images is far denser 
than than the HST images. 
In addition, their different perspective of a sun-target-observer angle of $\approx$15 deg, versus $<$2 deg from Earth, 
enables better constraints on the planes of their orbits \citep{2016Sci...351.0030W}. 
The combination of HST and LORRI images of the Pluto system thus provide the best available dataset to constrain the 
orbits of the small satellites. 

Unlike most small bodies in the Solar System, the small satellites of Pluto have the distinct advantage of being imaged up 
close by a spacecraft. 
Both Nix and Hydra were extensively imaged during the Pluto flyby, with image resolutions up to 0.3 km/pixel for Nix and 1.1 km/pixel for 
Hydra \citep{2016Sci...351.0030W}. 
Kerberos and Styx were discovered after the Pluto flyby imaging sequence had been finalized, so they could not be imaged as 
extensively, despite Styx being the small satellite that came closest to the spacecraft \citep{2016Sci...351.0030W}. 
Three placeholder imaging sequences had been left to be determined later; two of these were used for Kerberos, 
and one for Styx, with best resolutions of 3.1 and 4 km/pixel respectively \citep{2016Sci...351.0030W}. 
Nix was roughly equator-on as seen by New Horizons during the flyby, and has a fast rotation rate of 1.829$\pm$0.009 hours, leading to good coverage of its surface
\citep{2016Sci...351.0030W}. 
This was sufficient for \citet{2021porterChapter} to estimate its volume as equal to that of a sphere with a diameter of 36.5$\pm$0.5 km. 
Hydra rotates even faster with a period of 0.4295$\pm$0.0008 hours, but was seen as roughly pole-on from the spacecraft 
\citep{2016Sci...351.0030W}. 
These images, along with a constraining non-detection on the Hydra look-back image sequence, were
still sufficient for \citet{2021porterChapter} to constrain the volume of Hydra as equal to that of a sphere with diameter  
36.2$\pm$1 km. 
Detailed shape modeling was not possible for Styx and Kerberos with their limited low resolution images, 
but \citet{2021porterChapter} were able to constrain their equivalent spherical diameters as 
10.5$\pm$3 km for Styx and 12$\pm$3 km for Kerberos.

In this work, we combine the LORRI and HST astrometry of the small satellites of Pluto in order to determine their 
orbits and dynamical masses, and then combine those masses with the volumes in \citet{2021porterChapter} to estimate their densities.
Given the sizes of the small satellites relative to Pluto and Charon, the mutual perturbations of the small satellites 
on each other is very small. 
However, it is not zero. 
In particular, the perturbations of Nix and Hydra onto Styx and Kerberos are significant, 
and can (as we show) be used to constrain the masses of Nix and Hydra. 
Likewise, minimizing the perturbations of Styx and Kerberos on Nix and Hydra can provide upper limits on the 
masses of Styx and Kerberos (though this does not provide lower limits). 
We can therefore use the combined constraints from the HST and LORRI data to place reasonable limits on the masses of all four satellites.

Dynamically constraining the masses of the small satellites of Pluto is not a new technique. 
\citet{2008AJ....135..777T} was the first major effort in this direction, using the initial HST observations from 2006-2008, 
as well as historical data for the Pluto-Charon binary. 
However, only Nix and Hydra were known at the time and this was a very short dataset. 
\citet{2008AJ....135..777T} found dynamical masses (i.e. GM, the gravitational constant multiplied by the inertial mass) for 
Nix of 3.9$\pm$3.4$\times$10$^{-2}$ km$^3$/s$^2$ and Hydra of 2.1$\pm$4.2$\times$10$^{-2}$ km$^3$/s$^2$.
They further applied a Charon-like density of 1.63 g/cm$^3$, resulting in diameters of 88 km for Nix and 72 km for Hydra. 
These sizes are about twice as large as was later determined from the New Horizons resolved images 
\citep{2016Sci...351.0030W,2021porterChapter}, but were very close for the quality of the data at the time. 
However, these sizes assumed a Charon-like density, corresponding to that of a non-porous rock-ice object.  
This is a reasonable assumption for a body like Charon that has compacted itself into a nearly spherical shape
\citep{2017Icar..287..161B},
but not for bodies like Nix and Hydra which are roughly one tenth the diameter of Charon, and clearly non-spherical. 

\citet{2012ApJ...755...17Y} used the discovery of Kerberos to constrain the masses of Nix and Hydra.
This was logical, as Kerberos is sandwiched between Nix and Hydra, and appeared much dimmer and therefore likely smaller
(which was later verified by the New Horizons images).
However, only the initial discovery astrometry of Kerberos was available, providing a very short arc.
\citet{2012ApJ...755...17Y} found that the existence of Kerberos placed upper limits to the dynamical masses of Nix to be
$<$3$\times$10$^{-3}$ km$^3$/s$^2$ and Hydra to be $<$6$\times$10$^{-3}$ km$^3$/s$^2$.
They were not able to constrain the mass of Kerberos, and it was effectively a massless test particle, but they were
able to constrain the orbital eccentricity of Kerberos to be $<0.02$.

\citet{2015Icar..246..317B} published the first full mass and orbit solution for the Pluto system which included all four
small satellites.
Using the HST observations though 2012, they were able to constrain the dynamical masses to be 
3.0$\pm$2.7$\times$10$^{-3}$ km$^3$/s$^2$ for Nix, 3.2$\pm$2.8$\times$10$^{-3}$ km$^3$/s$^2$ for Hydra,
1.1$\pm$0.6$\times$10$^{-3}$ km$^3$/s$^2$ for Kerberos, and an upper limit of 
1.0$\times$10$^{-3}$ km$^3$/s$^2$ for Styx.
Notably, the mean mass for Nix in this solution was at the upper limit of Nix's predicted mass from \citet{2012ApJ...755...17Y},
though Hydra was well within the limit from \citet{2012ApJ...755...17Y}.
\citet{2015Icar..246..317B} also predicted a much larger mass for Kerberos than \citet{2012ApJ...755...17Y} found to
be long-term stable.
This high mass for Kerberos was likely a result of the high mass of Hydra in their solutions, which required a commensurately
higher mass for Kerberos to be more resistant to the perturbations of Hydra.
With Kerberos's post-flyby volume equivalent diameter of $\sim 12$ km \citep{2021porterChapter}, the Kerberos mass from \citet{2015Icar..246..317B} at face value would imply an
unreasonably high bulk density for Kerberos of 16 g/cm$^3$. 
However, their 1-sigma uncertainty on the mass of Kerberos was very large, $\sim 55\%$, so that lower masses and physically plausible densities were not excluded.

\citet{2019AJ....158...69K,2022AJ....163..238K} have investigated the long term dynamical stability of the Pluto small satellite system,
using the \citet{2015Icar..246..317B} states as initial conditions. 
\citet{2019AJ....158...69K} performed 1 Gyr simulations, and found that  the masses of Nix and Hydra were likely to be 
$\approx$10\% larger than the masses from \citet{2015Icar..246..317B}.
\citet{2022AJ....163..238K} extended this work to explore the possible masses of all four satellites. 
Using the rough sizes from \citet{2016Sci...351.0030W}, they constrained the densities of all four to be $\le$1.4 g/cm$^3$.
This is a more powerful long-term constraint than the simulations presented here, but is based on less data
(pre-2013 HST only), and is consistent with our results.

After the Pluto flyby, the JPL NAIF group made available an updated Pluto system solution named 
PLU055 
\footnote{\url{https://naif.jpl.nasa.gov/pub/naif/generic_kernels/spk/satellites/}; file: \texttt{plu055.cmt}}.
This solution added the initial LORRI astrometry from the New Horizons encounter to the solution from \citet{2015Icar..246..317B},
and refit the masses and orbits of Pluto, Charon, and the small satellites.
It was however only a best-fit solution, and did not provide any uncertainty analysis.
The masses for the small satellites in PLU055 are 2.15$\times$10$^{-3}$ km$^3$/s$^2$ for Nix,  
3.66$\times$10$^{-3}$ km$^3$/s$^2$ for Hydra, 
0.454$\times$10$^{-3}$ km$^3$/s$^2$ for Kerberos, and
2$\times$10$^{-20}$ km$^3$/s$^2$ for Styx, as Styx in their solution would attempt to have zero or even negative mass.
With the mean volumes from \citet{2021porterChapter}, these masses imply densities of 1.26 g/cm$^3$ for Nix, 
2.21 g/cm$^3$ for Hydra, and 7.5 g/cm$^3$ for Kerberos.
These were reasonable solutions for Nix and Hydra, but the high mass for Kerberos (and zero mass for Styx) still implied 
that the solution had not converged.
For our analysis, we used the PLU055 masses and states for Pluto and Charon without variation, and used the PLU055 
state vectors as initial conditions for the small satellites.

In this work, we extend and expand upon these prior mass and orbit estimates with two key innovations: 1) fully
reprocessed and self-consistent HST and LORRI astrometry for the small satellites, and 2) density constraints using the shape
models derived from the New Horizons resolved images.
The first innovation provides the highest quality dataset to fit the orbits of the small satellites, and 
the second ensures that the solutions produced are physically reasonable.
As a result, we are able to produce greatly improved estimates of the masses and densities of Nix and Hydra, as well as the best orbit solution for the entire system.
Our results show that Nix and Hydra are substantially less dense than Charon, and that the limited data for both Styx and Kerberos are consistent with their having similar densities to Nix and Hydra.
These densities, together with their bright, ice-rich surfaces, imply that Nix and Hydra are most likely more 
ice-rich than Charon with relatively low bulk porosities ($< 35$\%).
This appears generally consistent with the formation of the small satellites from debris ejected during a Charon-forming giant impact \citep{2011AJ....141...35C}. 

\section{Astrometric Analysis}

In order to fit the orbits and masses of the small satellites of Pluto, we needed astrometry of them. 
The only available sources for this astrometry are from images obtained by HST and New Horizons. 
The previously published 2005-2015 HST astrometry of the small satellites was all measured relative to Pluto 
\citep[e.g.][]{2015Icar..246..317B}, 
which is problematic due to Pluto's prominent albedo features creating significant center-of-light/center-of-mass offsets 
\citep{2012AJ....144...15B}. 
In addition, astrometry for the 2017-2019 HST observations (i.e. after the New Horizons flyby) has not been previously published. 
Since Pluto was in front of dense starfields for all of the HST images, we were able to register the HST images to stars in the Gaia DR2 
star catalog \citep{2018A&A...616A...2L}, and then perform absolute astrometry for the small satellites. 
Absolute astrometry is useful, as the New Horizons flyby allowed both the Pluto system heliocentic orbit and 
Pluto-Charon orbit to be determined to high precision \citep{2012AJ....144...15B,2015Sci...350.1815S}. 
The other dataset for small satellite astrometry is the New Horizons LORRI 4x4 images, which span the period from February 2015 to 
early July 2015. 
Preliminary astrometry of this dataset was used for the orbit/mass solution of PLU055, 
but that analysis was not registered to the Gaia DR2 star catalog, and only included the relatively deep \verb|U_HAZ| observations. 
Since then, the New Horizons mission has updated the pointing solutions for all of the LORRI 4x4 images to use Gaia DR2. 
We thus also set out to reanalyze the small satellite astrometry of the entire relevant LORRI 4x4 dataset. 
Our combined HST and LORRI astrometry is the best available dataset to fit the orbits and masses of the small satellites of Pluto.

\subsection{HST Astrometric Analysis}

\movetabledown=50mm
\begin{rotatetable}
\begin{deluxetable*}{lcccclcccccc}
    \tablecaption{ 
        Astrometry of the Small Satellites of Pluto as Observed by the Hubble Space Telescope (HST).
        \label{tab:hst}
    }
    \tablehead{
        \colhead{UTC Observation} &
        \colhead{HST} &
        \colhead{HST} &
        \colhead{HST} &
        \colhead{Exposure} &
        \colhead{Sat.} &
        \colhead{${\delta}RAcDE$} &
        \colhead{${\delta}DE$} &
        \colhead{${\sigma}RAcDE$} &
        \colhead{${\sigma}DE$} &
        \colhead{Corr.} &
        \colhead{SNR} \\
        \colhead{Midtime} &
        \colhead{Camera} &
        \colhead{Dataset} &
        \colhead{Filter} &
        \colhead{Time (s)} &
        \colhead{\tablenotemark{a}} &
        \colhead{(mas)\tablenotemark{b}} &
        \colhead{(mas)\tablenotemark{b}} &
        \colhead{(mas)\tablenotemark{c}} &
        \colhead{(mas)\tablenotemark{c}} &
        \colhead{Coef.\tablenotemark{d}} &
        \colhead{\tablenotemark{e}}
    } 
    \startdata
        2005-05-15T00:24:58.2 & ACS/WFC & j96o01 & F606W & 475.0 & K & 887.3 & -2489.6 & 7.1 & 7.1 & -0.005 & 15.3 \\
        2005-05-15T01:42:01.2 & ACS/WFC & j96o01 & F606W & 475.0 & N & -1132.5 & 1828.3 & 6.7 & 6.8 & -0.010 & 65.3 \\
        2005-05-15T01:42:01.2 & ACS/WFC & j96o01 & F606W & 475.0 & H & -1818.9 & -88.6 & 6.9 & 7.0 & -0.003 & 40.3 \\
        2005-05-18T03:10:56.2 & ACS/WFC & j96o02 & F606W & 475.0 & N & -163.8 & 2098.4 & 6.7 & 6.8 & 0.010 & 75.0 \\
        2005-05-18T03:31:55.2 & ACS/WFC & j96o02 & F606W & 475.0 & K & -125.3 & -2336.7 & 6.8 & 7.0 & 0.004 & 14.3 \\
        ... \\
        2019-08-03T13:15:03.4 & WFC3 & idwc51 & F350LP & 195.0 & H & -2088.8 & -990.8 & 20.8 & 20.9 & -0.018 & 55.1 \\
        2019-09-01T10:08:02.9 & WFC3 & idwc53 & F350LP & 200.0 & S & -266.7 & -1470.0 & 33.3 & 32.1 & -0.100 & 2.5 \\
        2019-09-01T10:08:02.9 & WFC3 & idwc53 & F350LP & 200.0 & N & -145.2 & 1788.6 & 14.0 & 15.0 & 0.021 & 40.2 \\
        2019-09-01T10:08:02.9 & WFC3 & idwc53 & F350LP & 200.0 & K & -2229.4 & 726.7 & 16.8 & 16.9 & -0.012 & 5.8 \\
        2019-09-01T10:08:02.9 & WFC3 & idwc53 & F350LP & 200.0 & H & 1364.8 & -2234.3 & 14.4 & 15.0 & -0.006 & 51.9 \\
    \enddata
    \tablenotetext{a}{S=Styx, N=Nix, K=Kerberos, H=Hydra}
    \tablenotemark{b}{RAcosDE and DE offsets of satellite relative to the Pluto system barycenter (milliarcseconds)}
    \tablenotemark{c}{1-$\sigma$ uncertainties for RAcosDE and DE offsets (milliarcseconds)}
    \tablenotemark{d}{Correlation Coefficient of RAcosDE and DE offsets}
    \tablenotemark{e}{Detected Signal-to-Noise Ratio of the satellite; observations with SNR$<$2 were excluded}
    \tablecomments{Table 1 is published in its entirety in the machine-readable format; 
    please see the MRT header for a full description.
          A portion is shown here for guidance regarding its form and content. 
          The images can be found in STScI MAST at \dataset[10.17909/zgqg-ym71]{http://dx.doi.org/10.17909/zgqg-ym71}.
          }
\end{deluxetable*}
\end{rotatetable}

The small satellites of Pluto have been observed with four different methods on HST. 
The earliest observations were two orbits of HST in 2005 with the Advanced Camera for Surveys (ACS) Wide Field Camera (WFC) 
and the F606W filter, in which Nix and Hydra were discovered (Program GO-10427). 
This was followed by two orbits in 2006 with the ACS High Resolution Channel (HRC), 
in the F606W and F435W filters (Program GO-10774). 
Then in 2007, the Pluto system was imaged with 16 HST orbits using the Wide Field and Planetary Camera 2 (WFPC2) 
using the F555W (most images), F814W, and F439W filters (Program GO-10786). 
ACS/HRC become inoperative in 2007, and in 2009, HST Servicing Mission 4 (the final flight of the Space Shuttle to HST) 
removed WFPC2 and replaced it with the Wide Field Camera 3 (WFC3). 
All HST images of the Pluto system from 2010 to 2019 were with WFC3, consisting of 107 separate HST orbits with the F350LP 
(98\% of images), F606W (25 images), and F845M (8 images) filters 
(Programs GO-11556, GO-12436, GO-12725, GO-12801, GO-12897, GO-13315, GO-13667, GO-15261, GO-15505).
We thus designed our astrometric analyses of the HST images around WFC3, and then adapted those techniques to 
ACS/WFC, ACS/HRC, and WFPC2.
The images can be found in STScI MAST at \dataset[10.17909/zgqg-ym71]{http://dx.doi.org/10.17909/zgqg-ym71}.

We based our analysis of the WFC3 images on the astrometric analysis of Arrokoth presented in \citet{2018AJ....156...20P}. 
That analysis used a pre-release version of the Gaia DR2 star catalog to fit the pointing of WFC3 images to $<$10 milliarcseconds, 
enabling high-precision astrometry and orbit analysis of Arrokoth to guide New Horizons to a flyby of that TNO. 
This was done by updating the pointing of the World Coordinate System (WCS) for the images to match Gaia stars. 
While the Arrokoth images used the full 2048x4096 UVIS2 apertures, all of the WFC3 observations were obtained with 512x512 
pixel subframes centered on Pluto, using either the UVIS1-C512B-SUB (GO-11556) or UVIS2-C512C-SUB (all others) apertures. 
The Pluto images thus only covered 3\% of the area of the sky as the Arrokoth images. 
However, the star density in the background of the Pluto images was very high, particularly 2011-2015 when most of this 
data was obtained. 
We were thus able to produce Gaia star-fit pointing solutions for 83\% of the WFC3 images of Pluto 2010-2019 with 
exposure times $\ge 30$ seconds, and 97\% of images with exposure times $\ge 120$ seconds. 
We started by using the initial pointing reported for each image to find and download the appropriate Gaia DR2 stars for each image. 
We used the same conditions for the stars as \citet{2018AJ....156...20P}, 
essentially that they had low uncertainty in both their proper motion ($<5$ mas/year) and parallax ($<1$ mas), 
and plenty of observations from Gaia ($>$80 good observations), both signs of a good solution for a single star 
which was not a close binary or errant minor planet. 
We corrected the stars for parallax and proper motion for each image. 
This was a small correction in most cases, as the Gaia DR2 epoch is J2015.5 \citep{2018A&A...616A...2L}, 
right in the middle of our observations, 
and most of the stars used for this analysis are sufficiently dim that they have very small parallax motion. 

We then created ``smear PSFs'', which replicated the effective star point-spread function (PSF) shapes in the images. 
Because HST was tracking on Pluto while the images were being taken, the stars moved over the course of the exposure, 
causing them to become streaked. 
This is not a linear process, as the apparent motion of Pluto relative to the stars changes over the course of a single HST orbit. 
The longer exposures ($> 120$ seconds) were particularly susceptible to this effect, but those were also the only exposures where 
Styx was visible, so getting them right was important. 
We created the smear PSFs by first finding the apparent motion of Pluto relative to the stars as seen by HST at 100 times over the 
course of the exposure. 
We used the spiceypy Python wrapper for JPL SPICE \citep{2020JOSS....5.2050A} combined with the 
JPL-supplied DE430 planetary solution and the JPL NAIF ``hst.bsp'' HST reconstructed orbital solution. 
We then found an appropriate Tiny Tim model PSF \citep{2011SPIE.8127E..0JK} 
for the exposure time and filter, shifted the Tiny Tim PSF to each of the 100 offsets from exposure mid-time, 
clipped their size to be 51x51 pixels, and then took the average to find the effective smeared PSF at the mid-time.

For the WFC3 images, we used the ``FLC'' data formats. 
These are unreprojected images similar to the older ``FLT'' formats, but with improved reduction of cosmic rays. 
We preferred the unreprojected FLC images over the Drizzled products (which have the geometric distortion removed through 
interpolation) as we reprojected the images to extract the small satellite astrometry, and we wanted to minimize the total 
amount of interpolation. 
We loaded the images, masking out any pixels that were identified in the data quality extension as being saturated or 
affected by either crosstalk or scattered light ghosts. 
We also masked out an area corresponding to a 30,000 km radius around Pluto-Charon barycenter. 
We found that this was a sufficient area to block out most of the scattered light from Pluto and Charon, 
while still showing plenty of stars. 
The number of stars varied greatly, depending on when the image was taken, with between 1 and 18 stars being used in the 
ultimate WCS solutions. 
Because we had confidence in the roll angle of the reported HST pointing, we were only fitting the right ascension/declination 
offset for the images, and thus could produce solutions with a single matched star (albeit at lower precision to many-star 
solutions). 
We extracted 51x51 pixel windows centered on each star, and used the smeared PSF to find a best fit solution for the star 
within the window. 
We used that best fit as an initial condition for a Markov Chain Monte Carlo (MCMC) probability distribution function (PDF) 
of the pixel positional uncertainty of the star, and saved the result in a database. 
We visually verified the match, and if it was good, it was saved as the first pass WCS solution. 
The predicted pointing for some of the older images (particularly 2010-2011) was sufficiently far off that the 
stars were outside of the 51x51 windows, so for those cases, we used an initial manual offset to roughly center the stars in the 
windows. 
17\% of the WFC3 images were marked as unusable at this point; they were generally either very low exposure times (less than 
60 seconds; meaning that not enough stars were visible) or were guide star failures that resulted in unusable images. 

To make the second-pass WFC3 WCS solutions, we then used the initial solution to extract new windows centered on the 
predicted location of the stars with the first-pass WCS, and refit them with the smeared PSFs. 
We used automatic outlier rejection to progressively drop the worst stars until the mean offset of the stars from the 
new windows was greater than 0.5 pixels relative to the first-pass (visually-vetted) solution. 
This was only needed in a few cases, but was useful to prevent a discrepant star from polluting the solution. 
We then used \textit{emcee} \citep{2013PASP..125..306F} 
to find a final PDF of the pointing offset for each image and save it in a database. 
The log-probability function for \textit{emcee} was based on both the measured uncertainty in the pixel position of the stars, 
as well as the star uncertainties from Gaia, which were generally much lower. 
The final pointing uncertainty for the WFC3 images varied from as low as 1 mas (for solutions with $> 10$ stars) to up 
11.8 mas (for solutions based on a single star). 
Finally, we visually vetted all of the WFC solutions by plotting the predicted location of all the Gaia stars in the field 
(including those not used in the solution) over the image and visually verifying that they matched. 
In total, we were able to produce good solutions for 97\% of the WFC3 images of Pluto with exposure times greater than 120 seconds.

The process to analyze the ACS and WFPC2 images was generally similar to the procedure for the WFC3 images, 
and reused much of that code. 
The ACS/WFC images from program GO-10427 were just two orbits with four images each. 
One of those per orbit was only 0.5 seconds long, and was discarded. 
The remaining three images per orbit had exposure times of 475 seconds, the F606W filter, and Pluto was placed in 
the gap between the two chips. 
We performed the same star-fitting procedure as for WFC3 on Chip 2 of ACS/WFC; 
the choice of this CCD was arbitrary, as there were plenty of stars on either chip, 
and the pointing offsets for both chips should be equal. 
Our ACS/WFC solutions used 269-284 stars, and had a resulting pointing uncertainty of ~1.2 mas. 
We again used a lightly modified version of the WFC3 code for the ACS/HRC data from program GO-10774. 
The images in program GO-10774 consisted of seven images with exposure times of 1 second (all these were discarded), 
one image at 3 seconds (also discarded), two images at 145 second exposures and the F606W filter, 
two images at 475 seconds and the F435W filter, and five images at 475 seconds and the F606W filter. 
The processing of these was almost identical to the subframed WFC3 images, though with FLT format images, 
as FLC is not available for ACS/HRC. 
The field of view of ACS/HRC is slightly larger than the 512 pixel apertures on WFC3, 
though with smaller and less sensitive pixels. 
Our WCS solutions for ACS/HRC thus used 8-14 stars, but had a pointing uncertainty of ~6 mas. 
We investigated using the ACS/HRC images of the Pluto system obtained in 2002 by program GO-9391, 
but they were of sufficiently short exposure time ($< 10$ seconds) that no stars were visible, 
and they were too faint to expect to see the small satellites anyway. 
The unique design of WFPC2 caused us to slightly change the design of our WCS fitting algorithm. 
Program GO-10786 was specifically designed as follow up the discovery of Nix and Hydra, and consists of 
14 HST orbits of observations in F555W and one orbit each in F439W, F675W, and F814W. 
Each F555W orbit had four exposures at 14 seconds (which were not used) and nine exposures at 100 seconds. 
F439W had five 350 second exposures, F675W had no exposures longer than 50 seconds and was discarded, 
and F814W had eight 140 second exposures. 
GO-10786 also imaged the Pluto system with NICMOS, but since those images are contemporaneous to the WFPC2 images 
and small in number, we did consider them. 
To fit these images, we first separately fit the stars on the three wide field cameras 
(Pluto was always on the narrow-field Planetary Camera), 
took the average of the three wide field solutions, and used that as initial conditions to solve the Planetary Camera pointing. 
This took full advantage of both the large number of stars in the wide field cameras enabling rapid unique solutions, 
and the superior angular resolution of the Planetary Camera enabling high-precision star fitting. 
With this technique, we were able to fit all of the WFPC2 images with exposure times longer than 50 seconds 
with between 6 and 36 stars, and final pointing uncertainties of 0.8-2.0 mas.

With the pointing solutions determined for all of the HST data, we next turned to minimizing the influence of scattered light from 
Pluto and Charon. 
This was compounded by the fact that Pluto is approximately 100 mas across from Earth, roughly 2.5 WFC3 pixels, 
while Charon has a rough angular diameter from Earth of 52 mas.  
For the 1521 WFC3 images in F350LP, we masked off the stars and Charon, reprojected the images to be centered on Pluto, 
and divided by the square of the distance to Pluto and by the estimated relative Pluto lightcurve based on reprojecting the Pluto 
albedo map from \citep{2017Icar..287..207B}. 
We stacked all of these reprojections to create a deep model of Pluto's effective PSF as seen by WFC3. 
We then reprojected the model Pluto back to the projections of the data, again correcting for distance and lightcurve. 
We then subtracted the Pluto model from the data, reprojected all the images to be centered on Charon, and stacked them. 
We could then reproject the Charon stack back to the data, add it to the Pluto model, 
and create combined Pluto-Charon models for all of the WFC3 F350LP data. 
We did not do this for the F606W and F814W filters, as there were not enough images to make usefully independent models. 
Likewise, we performed the same modeling on the 126 F555W images from WFPC2, but not the other filters, nor for any of the ACS data. 
Subtraction of the Pluto-Charon model was not always necessary to detect the satellites, but was helpful to maximize their 
signal-to-noise ratios, particularly for Styx and Kerberos.

Like the pointing solutions, we designed our astrometry around the WFC3 F350LP dataset, and then made adjustments for the rest of 
the HST data as appropriate. 
For each satellite and camera, we first used JPL SPICE and our WCS solutions to find which satellite was on which chip 
(necessary for ACS/WFC, but not for the other cameras), and which filter and exposure time the image had. 
We filtered the WFC3 images with exposure time; Nix and Hydra used images with 30 second exposures or longer, 
Kerberos 90 seconds or longer, and Styx 120 seconds or longer. 
For ACS, we used the same filtering, but did not look for Styx (V$\approx$27) at all, as Kerberos (V$\approx$26) 
was not clearly detectable in the images from either the WFC or HRC. 
Likewise, we did not search for either Styx or Kerberos in the WFPC2 images. 
We grouped the images by HST orbit and filter, so that if an orbit used two different filters, 
there would be two different groups per orbit. 
For each of these orbit/filter groups, we found a mean roll angle, subtracted the Pluto/Charon model for the cases we had it, 
and then reprojected all of the images of the group to be centered on the satellite and with the mean roll angle. 
We then stacked the images, and fit them with an appropriate Tiny Tim model PSF \citep{2011SPIE.8127E..0JK} with free parameters DN/s flux 
and x/y pixel offset. 
We made sure that the DN/s flux was larger than zero, but also smaller than twice the approximate flux of magnitude 
23 for Nix and Hydra, 26 for Kerberos, and 27 for Styx. 
This was to minimize the cases where the fitter optimized on a brighter star or noise spike rather than the satellite. 
We used the best fit with Tiny Tim as initial conditions for \textit{emcee} \citep{2013PASP..125..306F} 
to find a full flux/x/y PDF for the stacked orbit. 
Then, for each image, we used the stack x/y PDF cloud and the individual image's WCS to create an initial right ascension 
and declination PDF. 
We convolved this initial cloud with the uncertainty of the WCS solution and the RA/dec uncertainty appropriate for 
100 km of uncertainty in the location of Pluto's barycentre relative to HST. 
This  estimate is likely conservative, but only inflates the uncertainty by 4 mas, or one quarter of WFC3 pixel. 
Finally, the resulting mean values and covariance matrix of the satellite absolute astrometry was saved along with 
the appropriate metadata in a numpy file.
An example of the HST astrometry is shown in Table \ref{tab:hst}, 
the residuals are shown in the first pane of Figure \ref{fig:residuals},
and the full astrometry set is available as a Machine-Readable Table (MRT).

\subsection{New Horizons LORRI Astrometry Analysis}

We have also performed updated Gaia DR2-fit astronomic analyses of the New Horizons LORRI approach images of small satellites. 
The New Horizons Long Range Reconnaissance Imager (LORRI) is a f/12.6 Ritchey-Chretien telescope with a 20.8 cm aperture, 
paired with a back-illuminated unfiltered CCD with 1024x1024 illuminated pixels \citep{2008SSRv..140..189C}. 
The native pixel scale is 1.02 arcseconds per pixel. 
However, because the New Horizons spacecraft does not have reaction wheels and LORRI is hard mounted to the spacecraft, 
all pointing of LORRI is performed with New Horizons' thrusters. 
For any exposures longer than 0.1 second, the spacecraft's pointing will drift within a ``deadband'' of 3.8 arcseconds as the 
thrusters occasionally fire to maintain pointing at a target. 
All of the LORRI images used for this analysis were longer than 1 second, and most were 10 seconds long. 
These images therefore utilized LORRI's ``4x4'' mode, which performs 4x4 on-chip binning of the images, from 1024x1024 images with 
1.02 arcseconds per pixel to 256x256 images with 4.08 arcseconds per pixel. 
The resulting images contained the LORRI deadband within a pixel, and were one quarter the data volume, allowing for faster 
downlink from the spacecraft to Earth. 
All of the observations consisted of several images grouped into a ``REQID'', with at least five images per observation. 
Multiple images per observation not only increased the depth of the observation, but also allowed for filtering out of 
cosmic ray strikes, which were a significant source of noise on the images. 
The LORRI observations can be roughly grouped into \verb|OPNAV|, \verb|U_HAZ|, and \verb|SCI/SATPHOT| sequences. 
The \verb|OPNAV| images cover the time from when Nix and Hydra were first detectable by LORRI in February 2015 through to the flyby, 
and were obtained to be used in the optical navigation of New Horizons to Pluto \citep{2017SPIE10401E..0WC}.
They are therefore not as deep as the other observations, but are more frequent. 
The \verb|U_HAZ| or ``hazards'' images were seven observation groups obtained on approach to Pluto with the intention of 
searches for any additional satellites or rings that might potentially create a hazard to the spacecraft during the flyby 
\citep{2018Icar..301..155L}.
These observations were much longer and deeper than the \verb|OPNAVs|; indeed, some were so long that we broke them up for analysis, 
since the satellites moved a measurable amount over the observation sequence. 
The \verb|SCI| images were astrometric/photometric observations of the small satellites between the main \verb|U_HAZ| observations, 
and the \verb|SATPHOT| observations were a few sequences designed to obtain photometry of the small satellites between the 
\verb|U_HAZ| observations and the start of resolved observations of small satellites in unbinned 1x1 mode. 
Because the absolute pointing knowledge of the New Horizons spacecraft is lower than the precision of the images, all of the LORRI 
4x4 images had to have their pointing updated by registering the stars in the images to the Gaia DR2 star catalog 
\citep{2018A&A...616A...2L}. 
During the New Horizons flyby, this was performed using the older UCAC4 star catalog \citep{2013AJ....145...44Z}, 
and that was the basis for the LORRI astrometry used for PLU055.
The Gaia DR2 catalog was released after the Pluto flyby in 2015, but before the Arrokoth flyby in 2019. 
In preparation for the Arrokoth flyby, the New Horizons project refit all the Pluto flyby 4x4 images with Gaia DR2, and those refit 
images are what we used for our analysis. 
Because the star field behind Pluto was very dense, and the field of view of LORRI much larger than WFC3, there were typically 
several hundred stars that could be used to fit the pointing of the LORRI images. 
These large number of control points meant that the pointing of each LORRI 4x4 image could be reliably determined to a precision of 
$<0.025$ 4x4 pixels, or $<0.1$ arcseconds. 
The physical resolution of the LORRI 4x4 images starts at roughly 3000 km/pixel in February 2015 and ends at roughly 300 km/pixel 
in early July 2015. 
We did not consider the LORRI 1x1 imagery of the small satellites in the astrometric analysis, as they only extend the LORRI data 
coverage by one week (i.e. the week before the July 14, 2015 flyby), and are a much more difficult dataset to perform astrometry on 
due to the resolved shapes of the satellites. 
Several HST observations of the Pluto system were performed during the approach of the New Horizons spacecraft to Pluto, 
making them contemporaneous with the New Horizons LORRI observations. 
This provided a unique and powerful constraint on the location of the small satellites at the time of the flyby, 
and is why we set the epoch of our orbit determination to be the time of the flyby.

\movetabledown=60mm
\begin{rotatetable}
\begin{deluxetable*}{lcccccccccccc}
    \tablecaption{ 
        Astrometry of the Small Satellites of Pluto as Observed by the New Horizons LORRI Camera.
        \label{tab:lorri}
    }
    \tablehead{
        \colhead{UTC Observation} &
        \colhead{Observation} &
        \colhead{Texp} &
        \multicolumn{3}{c}{S/C Pos.\tablenotemark{a} (au)} &
        \colhead{Sat.} &
        \colhead{${\delta}RAcDE$} &
        \colhead{${\delta}DE$} &
        \colhead{${\sigma}RAcDE$} &
        \colhead{${\sigma}DE$} &
        \colhead{Corr.} &
        \colhead{SNR} 
        \\
        \colhead{Midtime} &
        \colhead{Name} &
        \colhead{(s)} &
        \colhead{x} &
        \colhead{y} &
        \colhead{z} &
        \colhead{\tablenotemark{b}} &
        \colhead{(as)\tablenotemark{c}} &
        \colhead{(as)\tablenotemark{c}} &
        \colhead{(as)\tablenotemark{d}} &
        \colhead{(as)\tablenotemark{d}} &
        \colhead{Coef.\tablenotemark{e}} &
        \colhead{\tablenotemark{f}}
    } 
    \startdata
        2015-01-31T01:43:24.7 & \texttt{NAV\_NONCRIT\_031} & 9.9 & 0.0151 & -1.2665 & -0.3308 & H & -23.4 & -46.8 & 0.9 & 1.3 & -0.128 & 19.5 \\
        2015-02-02T01:25:24.7 & \texttt{NAV\_NONCRIT\_033} & 9.9 & 0.0149 & -1.2512 & -0.3268 & N & -36.8 & -1.4 & 1.3 & 1.4 & -0.111 & 15.7 \\
        2015-02-10T18:00:24.7 & \texttt{NAV\_NONCRIT\_041} & 9.9 & 0.0141 & -1.1842 & -0.3093 & N & 12.4 & 43.6 & 0.9 & 1.0 & -0.008 & 18.6 \\
        2015-02-15T01:37:24.7 & \texttt{NAV\_NONCRIT\_046} & 9.9 & 0.0137 & -1.1510 & -0.3006 & H & -10.4 & 71.7 & 0.4 & 0.4 & 0.093 & 26.2 \\
        2015-02-23T20:00:24.7 & \texttt{NAV\_CRIT\_054} & 9.9 & 0.0129 & -1.0834 & -0.2830 & H & 55.8 & 3.9 & 0.4 & 0.4 & 0.068 & 29.5 \\
        ... \\
        2015-07-03T04:28:00.9 & \texttt{SATPHOT\_9\_01} & 0.8 & 0.0009 & -0.0871 & -0.0227 & N & -543.5 & 322.0 & 3.0 & 3.044 & 0.005 & 239.5 \\
        2015-07-03T04:28:00.9 & \texttt{SATPHOT\_9\_01} & 0.8 & 0.0009 & -0.0871 & -0.0227 & H & -371.3 & -643.6 & 3.0 & 3.0 & 0.005 & 285.5 \\
        2015-07-03T04:29:15.9 & \texttt{SATPHOT\_9\_02} & 0.8 & 0.0009 & -0.0871 & -0.0227 & N & -543.5 & 322.2 & 3.0 & 3.0 & 0.011 & 212.1 \\
        2015-07-03T04:30:30.9 & \texttt{SATPHOT\_9\_03} & 0.8 & 0.0009 & -0.0871 & -0.0227 & S & 221.4 & 459.1 & 3.0 & 3.0 & 0.015 & 37.0 \\
        2015-07-03T04:31:45.9 & \texttt{SATPHOT\_9\_04} & 0.8 & 0.0009 & -0.0871 & -0.0227 & S & 221.6 & 459.1 & 3.0 & 3.0 & 0.006 & 34.2 \\
    \enddata
    \tablenotetext{a}{Position of the New Horizons spacecraft relative to the Pluto system barycenter in ICRF}
    \tablenotetext{b}{S=Styx, N=Nix, K=Kerberos, H=Hydra}
    \tablenotemark{c}{RAcosDE and DE offsets of satellite relative to the Pluto system barycenter (arcseconds)}
    \tablenotemark{d}{1-$\sigma$ uncertainties for RAcosDE and DE offsets (arcseconds)}
    \tablenotemark{e}{Correlation Coefficient of RAcosDE and DE offsets}
    \tablenotemark{f}{Detected Signal-to-Noise Ratio of the satellite; observations with SNR$<$2 were excluded}
    \tablecomments{Table 2 is published in its entirety in the machine-readable format; 
    please see the MRT header for a full description.
          A portion is shown here for guidance regarding its form and content.}
\end{deluxetable*}
\end{rotatetable}

Like for the HST images, a major impediment to measuring the locations of the small satellites was scattered light 
from Pluto and Charon. 
We created rendered model Pluto and Charon images, using the albedo maps, and convolved them with the LORRI PSF. 
In addition, the background stars in the LORRI images were much more of an impediment to detecting the small satellites, 
due to the low angular resolution of the LORRI 4x4s. 
In 2013 and 2014, LORRI imaged the area of the sky that the Pluto would appear in during the approach to flyby. 
We were thus able to build up a deep star background field covering the area imaged in the Pluto approach 4x4s. 
We reprojected the deep background stack to the frame of each individual image, and added the star background models 
and rendered Pluto/Charon models together to produce a model image of all the features in the images that were not 
the small satellites. 
In practice, after subtracting the models from the data, the remaining images contained the small satellites, 
cosmic rays, small scattered light artifacts, and any variable stars that were of a different brightness than they 
were when the background data was obtained. 
Our stacking process was able to minimize the cosmic rays and scattered light, but variable stars did cause a 
few astrometric points to be discarded.

Once all of the images were subtracted, we then processed the small satellites separately. 
We first examined each image to see if the satellite was on the field (not a given for the later LORRI observations) and then made a 
list of images and their associated REQID (observation sequence ID). 
For each REQID, we determined the mean roll angle of the observation; early \verb|OPNAV| observations rolled the spacecraft 
over the course of the observation, while the \verb|U_HAZ| and science observations used one or two roll angles for the full 
observation. 
We then reprojected all of the subtracted images for the REQID into new images that were centered on the predicted location of the 
satellite, had the mean roll angle, and were scaled up by a factor of four, effectively returning them to the original LORRI 
resolution. 
These reprojected images were then median-combined to produce a stacked image of the satellite. 
We also used the Gaia DR2 star catalog to find 30 fiducial stars in the images; 
we reprojected the image to be centered on each of those stars, and stacked them to produce an empirical model of the PSF of that 
image. 
The PSF of the images would vary due to the different thruster firings during each exposure. 
We stacked all of the model PSFs for a REQID in the same way as the reprojected images of the satellite to produce a model of the 
PSF for the stacked image. 
We fit the stacked PSF with a two dimensional normal distribution to produce a parametric model of the stacked PSF. 
We also performed the same reprojection and stacking operation on the model star/Pluto images, 
and used the result to identify any areas that had saturated (or nearly saturated) background stars that would not have 
subtracted cleanly, and thus could produce spurious results. 
We then fit the parametric model PSF to the data, and visually inspected the result. 
This was particularly important for Styx and Kerberos to make sure that the best-fit of the PSF was fitting the satellite 
and not an unsubtracted variable star or a residual noise spike from a very bright star. 
For all the images that were visually confirmed, we then used \textit{emcee} \citep{2013PASP..125..306F} 
to find the MCMC uncertainty distribution of the flux and x/y pixel offset for the stacked image. 
If the signal-to-noise ratio of the flux from \textit{emcee} was less than 3.0, we discarded the image and moved on. 
We then looked to see if the REQID consisted of multiple of pointings; 
this was typical of the \verb|U_HAZ| and \verb|SATPHOT| observations. 
If the REQID had multiple pointings, we produced an astrometric observation for each one, and only one observation otherwise. 
This was important because the satellites moved a considerable distance over the course of some of the \verb|U_HAZ| observations. 
We converted the pixel positions from the MCMC cloud to a astronomic right ascension and declination cloud with the image's WCS. 
We convolved this initial RA/Dec cloud with the uncertainty in the WCS ($<$0.1 arcseconds), as well as the angular uncertainty 
corresponding to 200 km at the distance from New Horizons to the targets. 
This is to cover the uncertainty in the location of the spacecraft relative to the Pluto system barycenter. 
Finally we saved the results in the same format as the HST data.
An example of the LORRI astrometry is shown in Table \ref{tab:lorri}, 
the residuals are shown in the second pane of Figure \ref{fig:residuals},
and the full astrometry set is available as an MRT.

\section{Mass and Orbit Fitting}

To simulate the Pluto system, we used a high-order Runge-Kutta-Nystom adaptive timestep integrator
\citep{10.1145/62038.69650}, 
written in C++ for speed, and with a Python interface \citep[PyNBody;][]{PyNBody}.
This code was originally developed for the New Horizons hazards analysis \citep{2018Icar..301..155L}, 
and has been verified through extensive testing on both the Pluto system and heliocentric orbit determination, 
notably for predicting the orbit and occultations of Arrokoth \citep{2018AJ....156...20P,2020AJ....159..130B}. 
We used the base integrator to set up a system comprising Pluto, Charon, and the four small satellites. 
Following \citet{2012ApJ...755...17Y},
we did not include solar perturbations, since those perturbations are very small compared to the interactions of the 
bodies in the Pluto system with each other. 
We parameterized the system with seven parameters for each of the four small satellites: 
the natural logarithm of the mass of the satellite, and the position and velocity of the satellite relative to the 
Pluto system barycenter at an epoch. 
We used the logarithm of the mass of the satellite to prevent the masses from becoming negative or zero. 
The cartesian position and velocity of the small satellites was in the 
International Celestial Reference Frame \citep[ICRF;][]{2009ITN....35....1M}, 
the same frame used for the Gaia DR2 star positions \citep{2018A&A...616A...2L}, 
and therefore the best comparison with our Gaia-fit astrometry. 
We chose the epoch for the satellite fitting as being shortly after the New Horizons Pluto flyby date, 
as that was the largest concentration of data, with observations from both New Horizons LORRI and HST happening simultaneously. 
This minimized the propagation from the epoch to the time of greatest constraint, 
but did mean that for each error estimation call we had to propagate both backwards and forwards from the epoch. 
The timebase used for the integrations was JPL Ephermeris Time, seconds in Barycentric Dynamical Time (TDB) since the J2000.0 epoch. 
This made it easy to interface the results of the simulations with the JPL SPICE library. 
We performed all the geometrical transformations from Pluto-system-barycentric states to astrometric 
Right Ascension/Declination as seen by either New Horizons or HST using the \textit{spiceypy} Python wrapper for 
JPL SPICE \citep{2020JOSS....5.2050A} 
and the JPL DE430 planetary ephemeris for the location of the Pluto system barycenter \citep{2014IPNPR.196C...1F}. 
We used the most recent New Horizons reconstructed kernel for the location of 
New Horizons (OD122
\footnote{\url{https://naif.jpl.nasa.gov/pub/naif/pds/data/nh-j_p_ss-spice-6-v1.0/nhsp_1000/}})
and the JPL NAIF kernel for HST's location relative to the 
Earth\footnote{\url{https://naif.jpl.nasa.gov/pub/naif/HST/kernels/spk/}; file: \texttt{hst.bsp}}.
We assumed that the target-observer light-time for all the satellites was the same as the light-time to Pluto-system barycenter, 
and saved this for all the observations. 
We could then determine a list of light-time-corrected target times for the integrator, removing any duplicates due to 
more than one satellite being observed at the same time (as was typically the case, except for the closest New Horizons images). 
To assess the error or probability of a given state, we would start by passing it to the Pluto system class, 
which would create a Pluto system with the masses and states of Pluto and Charon as defined by PLU055, 
the most recent Pluto system solution released by JPL, and then add the small satellites with the prescribed masses and states. 
The integrator would then integrate the system to the light-corrected times for all the observations, 
and pass back the resulting positions relative to the Pluto barycenter. 
We could then add the light-time-corrected vector from the observer to the Pluto system barycenter to obtain the observer-target 
vectors for all of the satellites, and therefore their RA/Dec astrometric positions. 
As noted above, we used absolute astrometry for all of our simulations, rather than astrometry relative to Pluto or Charon, 
to remove the affects of center-of-light/center-of-mass offsets inherent to those bodies. 

We compared the predicted positions of the small satellites to the observations to compute a $\chi^2$ for the solution. 
First, we performed sanity checks on the given state, and if the predicted density of any of the satellites is beyond the specified limits, 
that solution is rejected. 
We set the acceptable density range\footnote{Bounding the MCMC solutions was necessary to prevent the code from spending too 
much time exploring unrealistic solutions (e.g., with negative masses or densities greater than expected for small outer solar system bodies).  
With much more extensive astrometry, it would be possible to let the MCMC explore the parameter space without such bounds and 
then reject unphysical solutions based on a $\chi^2$ analysis.  
However, with current data and computational capabilities this was not feasible.}
to be 0.1-3.0 g/m$^3$, based on the volumes given in \citet{2021porterChapter} and plausible values for small outer solar system bodies.
In addition, we ran a stability check for each solution where we integrate the Pluto system backwards in time for a 
given number of years, periodically checking to make sure that each satellite's orbit's out-of-plane angular momentum component 
$\sqrt{1-e^2}{\times}cos(I)$, 
where $e$ is eccentricity of the satellite's orbit relative to the Pluto system barycenter and 
$I$ is the inclination of the satellite's orbit to the Pluto-Charon orbit,
 is less than $10^{-3}$. 
Styx is the most highly perturbed, and most likely to violate this condition.
For the solutions presented here, we ran this stability constraint for just 10 year prior to the epoch,
and then re-ran it for 1000 years afterwards to further verify the stability of the orbits.
None of the solutions that met our stability criteria for 10 year violated it for 1000 years.
Because 28 parameters can be unwieldy to solve, we have the ability to mask off any of the parameters. 
In practice, we generally have used this to hold the masses of the small satellites constant while the state vectors are initially 
solved, before allowing the mass and states to both vary freely. 

With this $\chi^2$ function and its density and stability priors, we could then construct a log probability function
for the mass and state vector solution suitable for use with the Markov Chain Monte Carlo (MCMC) package
\textit{emcee} \citep{2013PASP..125..306F}.
MCMC estimation is useful for the problem of the orbits and masses of the small satellites of Pluto, as all of the
parameters are dependent on each other in complicated ways.
A slight change to the mass of Nix, for example, would require a change to the mass of Hydra (since they perturb each other),
and to the state vectors of Styx and Kerberos (since they are heavily perturbed by Nix).
MCMC uses a group of ``walkers'' (300 in the case of our mass/orbit solutions) which jump through a series of states that
are equally likely in probability space.
Once the MCMC run has sufficiently evolved away from its initial conditions (``burned-in''), the paths of the walkers through
probability space represent a discretely-sampled version of the probability distribution function (PDF) for the 
given parameter space and log probability function.
The \textit{emcee} package provides a facilities to both check that a PDF is fully converged, and to restart a MCMC chain to 
push it closer to convergence, and we make use of both of these in producing our final uncertainties.

Initially, we ran several solutions where we held the masses of the satellites constant, found best-fit state
vectors for that combination of masses, and then ran \textit{emcee} with the states held constant and the
masses allowed to vary. 
We started with the states and masses from the PLU055 solution. 
This solution has somewhat high densities with the most recent volumes for Nix and Hydra (1.26 g/cm$^3$ and 2.21 g/cm$^3$ 
respectively), a very high density for Kerberos (7.5 g/cm$^3$), and effectively zero mass for Styx; 
the latter two features indicate a lack of sensitivity to the tiny masses of both Kerberos and Styx in the fit.
We disabled our density constraint for this case, as both Kerberos and Styx would be in violation of it.
We found that while this solution worked well around the time of the Pluto flyby, it was not a good solution for the pre-2012 HST astrometry, 
nor the post-flyby HST astrometry. 
Nix was particularly notable for this, as its reduced $\chi^2$ relative to PLU055 
for the WFC3 data was 0.202, while it was 13.5 for the ACS/WFPC2 dataset.
Allowing the solution to fit all of the data, and with our stability restriction running 1000 years into the past, 
we found the mass of Hydra was trending downwards, though bringing Kerberos' mass up along with it, leading 
to even more unrealistic densities. 
Seeing this, we enabled the 0.1-3.0 g/cm$^3$ density constraint, and 
started a second solution set with the PLU055 initial positions and the masses of all of the small satellites 
set such that their densities were 1.0 g/cm$^3$ with the nominal volumes \citep{2021porterChapter}. 
This produced a strong solution where the masses (and thus densities) of Nix and Hydra are similar, 
and strongly correlated to each other. 
This appeared to be a good and consistent dataset, but we also tried a third case with initial densities of 0.5 kg/cm$^3$. 
The 0.5 g/cm$^3$ solution quickly evolved into a ~0.3 g/cm$^3$ solution, which also seemed consistent.
The low-density case also produced similar masses for Nix and Hydra; because they perturb each other,
one cannot be much higher mass than the other, and (since their volumes are similar) therefore they must 
have similar densities.

\begin{figure*}
    \plotone{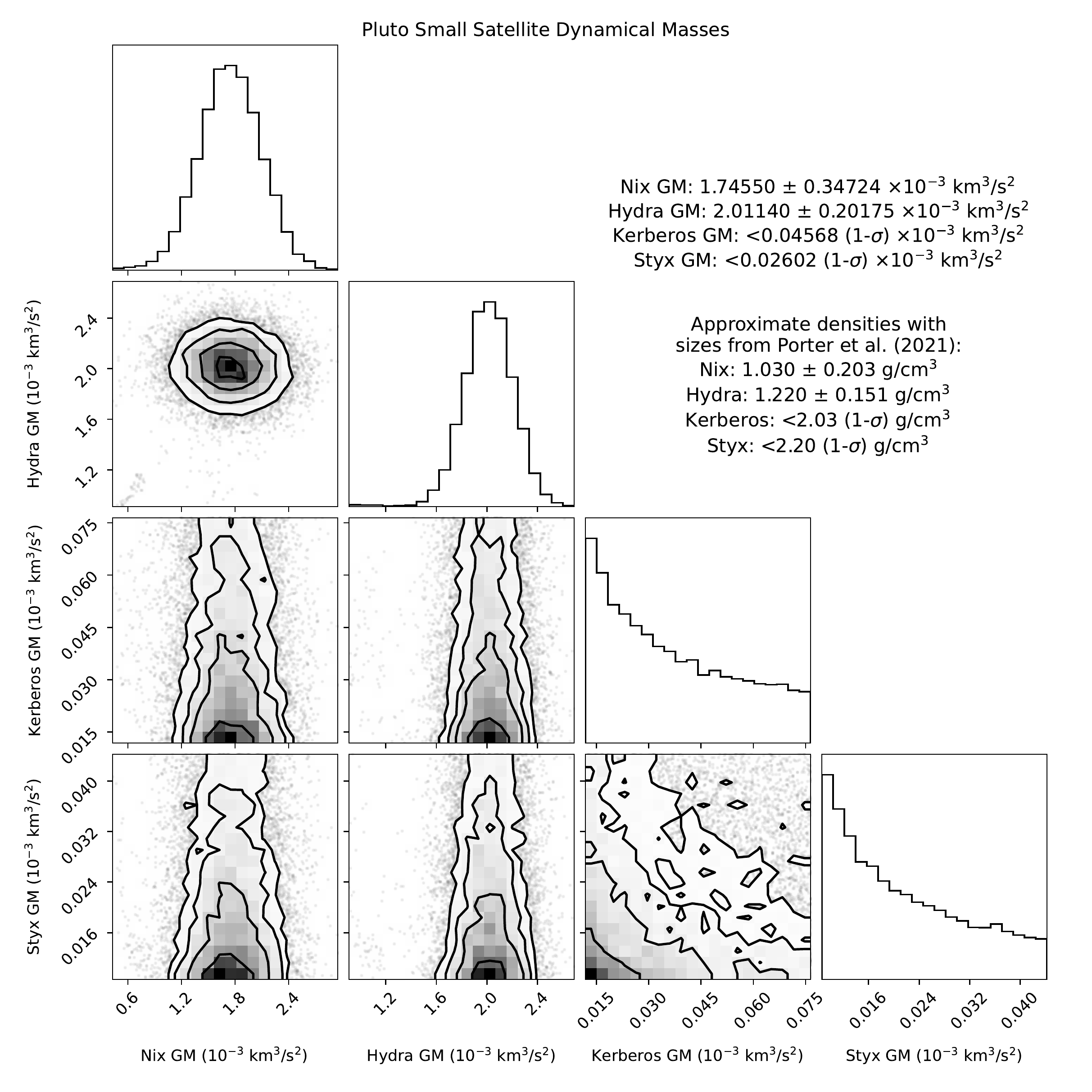}
    \caption{
        Corner plot \citep{2016JOSS....1...24F} of the dynamical masses of the small satellites of Pluto derived from 
        the family of solutions produced by the Markov Chain Monte Carlo model.
        Histograms on top show the probability distributions for the masses of each object, while contour plots
        show the probability covariance between the masses. 
        In the contour plots, all individual MCMC solutions are overplotted, with each solution being equally likely.  
        Darker [lighter] regions have a higher [lower] density of such solutions, and thus show regions of higher [lower] overall likelihood.  
        Contour lines for solution density show the shape of the covariance between the parameters.  
        The mean of all MCMC solutions provides the expectation value of the solution set, 
        and the standard deviation provides the uncertainty, with values given in the upper right. 
        Both Nix and Hydra have well-defined masses, and by combining these with the volumes from \citet{2021porterChapter},
        we find that Nix and Hydra have densities close to 1.1 g/cm$^3$, substantially lower than the densities of Pluto and Charon. 
        The Kerberos and Styx masses remain poorly constrained, with 1-sigma upper limits for their masses given above.
    }
    \label{fig:corner}
\end{figure*}

\begin{figure*}
    \plottwo{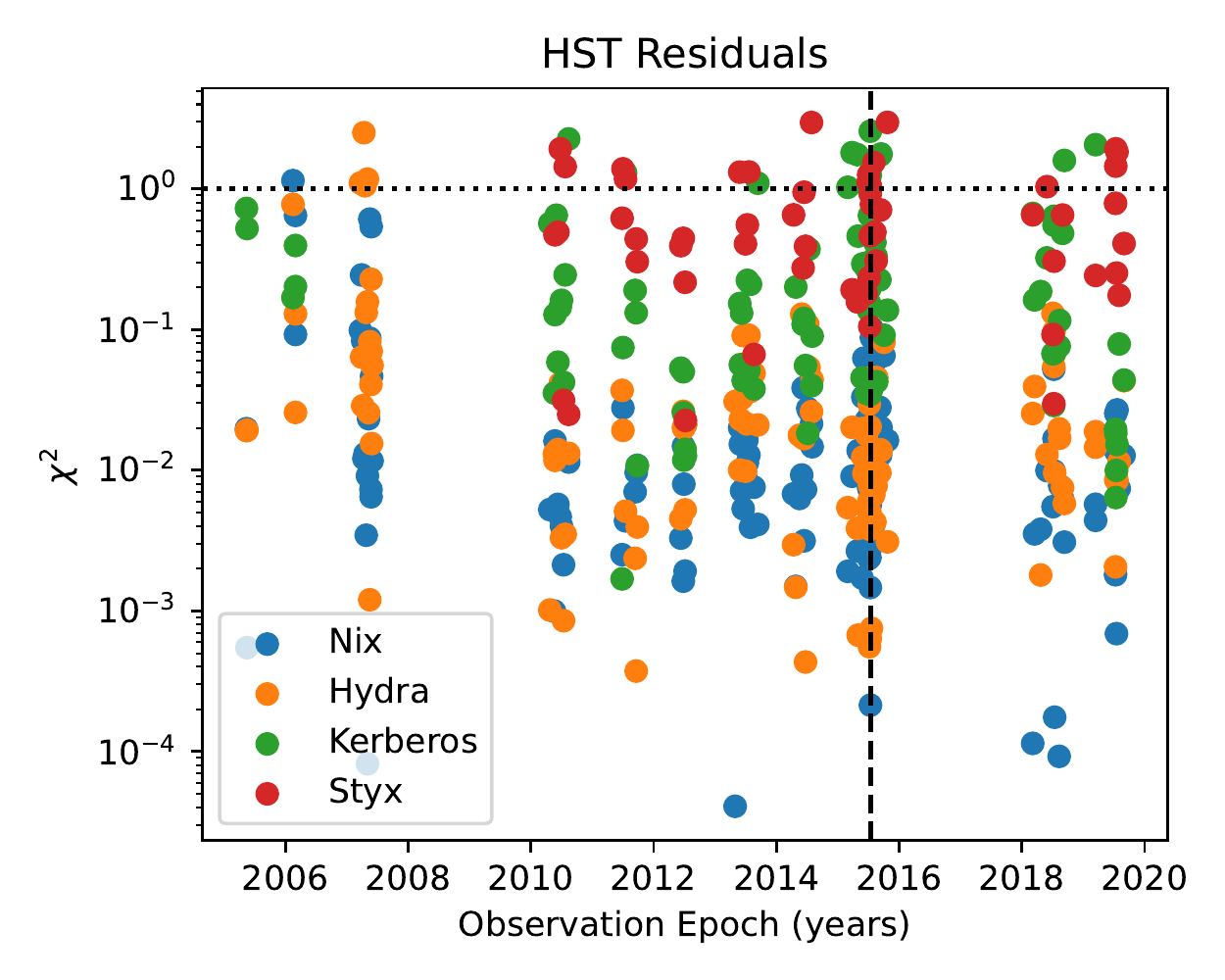}{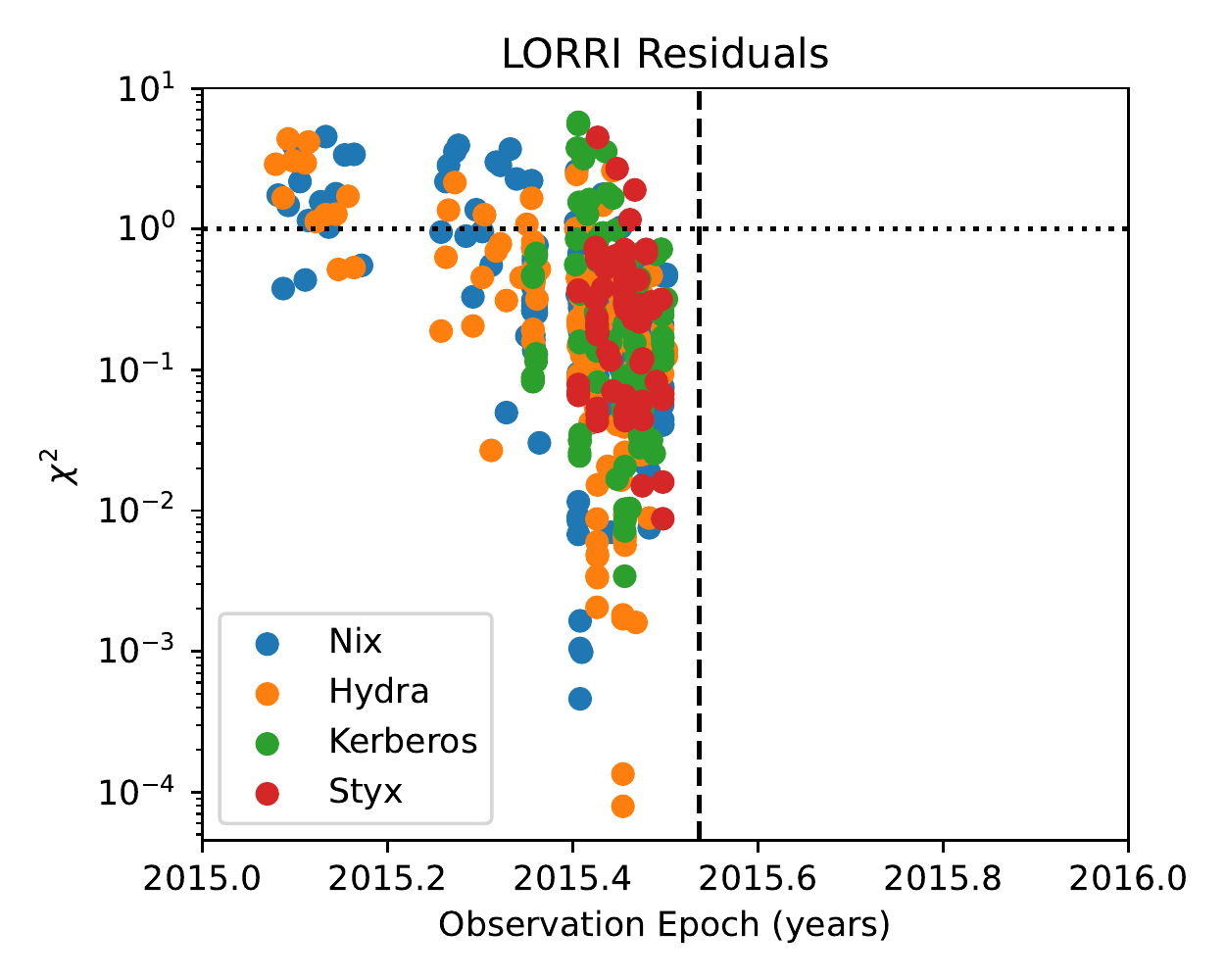}
    \caption{
        Residuals for the nominal solution given in Table \ref{tab:states} versus the HST (left) and New Horizons datasets (right).
        The horizonal dotted line is at $\chi^2$=1, and the vertical dashed line is the New Horizons flyby.
        Note that the highest residuals are for Styx and Kerberos in all the HST data, and for Nix and Hydra in the early 
        LORRI observations; these correspond to low SNR observations which often have excessive error beyond what would be expected from the 
        resolution of the camera due to faint background sources (Pluto was in a dense star field for all these observations).
        By combining the HST and New Horizons datasets, we were able to combine the best Nix and Hydra constraints from HST with
        the strongest Styx and Kerberos constraints from New Horizons. 
        The reduced $\chi^2$ for each body for this particular solution, 
        averaging across all the data sets (including those with low signal-to-noise), 
        are 0.376 for Nix, 0.331 for Hydra, 0.596 for Kerberos, and 0.676 for Styx; see the text for more details.
    }
    \label{fig:residuals}
\end{figure*}

We finally combined our low-density ($\approx$0.3 g/cm$^3$) and high-density cases ($\approx$1.0 g/cm$^3$) together
and allowed all free parameters to vary, both masses and states.
This took over a week to run on a 32-core desktop computer, and the results are shown in Figure \ref{fig:corner} and Table \ref{tab:states}.  
Residuals for the nominal solution given in Table \ref{tab:states} are shown in Figure \ref{fig:residuals}.

We find that the dynamical mass of Nix is 1.74$\pm$0.35$\times$10$^{-3}$ km$^3$/s$^2$ (2.60$\pm$0.52$\times$10$^{16}$ kg) and 
the dynamical mass of Hydra is 2.01$\pm$0.20$\times$10$^{-3}$ km$^3$/s$^2$ (3.01$\pm$0.30$\times$10$^{16}$ kg). 
Combining the volumes and volume uncertainties from \citet{2021porterChapter} with these masses produces densities of
1.031$\pm$0.204 g/cm$^3$ for Nix and
1.220$\pm$0.150 g/cm$^3$ for Hydra. 
The positional uncertainty for Nix and Hydra at the epoch was $\approx$4 km, $\approx$8 km for Kerberos, and $\approx$11 km
for Styx (or roughly the size of Styx).
In addition to the primary density solution peak at $\approx1.1$ g/cm$^3$ for Nix and Hydra, there is a much less likely secondary
solution class where Nix roughly has a mass of 0.6$\times$10$^{-3}$ km$^3$/s$^2$ and Hydra has a mass of 
1.0$\times$10$^{-3}$ km$^3$/s$^2$, corresponding to densities of 0.4 g/cm$^3$ for Nix and 0.6 g/cm$^3$ for Hydra.
This can be seen as a small number of faint points in the lower left corner of the Nix/Hydra covariance plot 
and is the remnant of our earlier low density solution, which is on the far edge of the full combined
probability distribution; these only accounted for 87 of 11400 cases (0.8\%) and thus had 
much lower overall likelihood than the main solutions near 1.1 g/cm$^3$.

\begin{deluxetable*}{c|cccccc}
    \tablecaption{ 
        ICRF state vectors of the small satellites of Pluto, relative to the Pluto system barycenter.
        The Pluto and Charon states are from the JPL PLU055 solution.
        \label{tab:states}
    }
    \tablehead{
        \multicolumn{7}{c}{Epoch: 2015-07-15 23:59:59.816 UTC, 2015-07-16 00:01:08.000 TDB, 490276868.000 ET}
        \\
        & \colhead{$x$ (km)} & \colhead{$y$ (km)} & \colhead{$z$ (km)} & 
        \colhead{$v_x$ (km/s)} & \colhead{$v_y$ (km/s)} & \colhead{$v_z$ (km/s)}
    }
    \startdata
        Pluto & 67.8 & -253.8 & -2118.2 & -0.017843 & -0.016437 & 0.001400 \\
        Charon & -554.9 & 2076.5 & 17330.3 & 0.145990 & 0.134489 & -0.011457 \\
        \hline
        Styx & -30168.4 & -28721.6 & -4092.4 & -0.010427 & 0.012730 & 0.150385 \\
        & $\pm$6.0 & $\pm$6.1 & $\pm$6.3 & $\pm$0.000025 & $\pm$0.000033 & $\pm$0.000014 \\
        Nix & 34094.4 & 29302.6 & -16558.3 & -0.012886 & -0.031895 & -0.132617 \\
        & $\pm$1.9 & $\pm$2.2 & $\pm$2.1 & $\pm$0.000006 & $\pm$0.000006 & $\pm$0.000003 \\
        Kerberos & -28868.4 & -19795.9 & 47919.3 & 0.087782 & 0.094671 & 0.076441 \\
        & $\pm$4.2 & $\pm$5.0 & $\pm$3.9 & $\pm$0.000009 & $\pm$0.000009 & $\pm$0.000008 \\
        Hydra & -22621.1 & -11900.4 & 61826.5 & 0.096768 & 0.097875 & 0.045197 \\
        & $\pm$2.1 & $\pm$3.2 & $\pm$1.2 & $\pm$0.000004 & $\pm$0.000004 & $\pm$0.000004 \\
    \enddata
\end{deluxetable*}
We estimate 1-$\sigma$ upper limits for the dynamical masses of 
Styx to be $<$0.03$\times$10$^{-3}$ km$^3$/s$^2$ ($<$5$\times$10$^{14}$ kg)
and Kerberos to be $<$0.05$\times$10$^{-3}$ km$^3$/s$^2$ ($<$8$\times$10$^{14}$ kg); 
however, the masses of these tiny satellites remain poorly constrained.
These 1-$\sigma$ upper mass limits, in combination with volume estimates from \citet{2021porterChapter}, imply densities of $<$2.0 g/cm$^3$ for Kerberos and 
$<$2.1 g/cm$^3$ for Styx; these densities, like their masses, are highly uncertain.

We ran the solution cloud in Figure \ref{fig:corner} back in time for 1000 years, recording the periods, eccentricities, and inclinations 
of all of the orbits every 0.1 years.
The resulting mean osculating orbital elements are shown in Table \ref{tab:elts},
and the residuals in Figure \ref{fig:residuals}.
These mean elements provide some insight into the quality and uncertainty of the orbit solution in Table \ref{tab:states},
and provide comparisons with previously published solutions.
However, we caution that these results should not be overinterpreted, as the long term circumbinary dynamics of Pluto system are complex,
and we could not capture any long term evolution of the orbits. 
See \citet{2015Natur.522...45S,2020AJ....159..277W,2021AJ....161...25B} for further discussion of the long term dynamics of Pluto system.

Figure \ref{fig:residuals} shows the residuals for the mean solution in Table \ref{tab:elts}
versus the HST and LORRI data.
The reduced $\chi^2$ for the HST data is: 0.053 for Nix, 0.102 for Hydra, 
0.534 for Kerberos, and 1.566 for Styx; 
for LORRI it is 0.643 Nix, 0.576 for Hydra, 0.841 for Kerberos, and 0.613 for Styx. 
The only dataset that was not overfit was thus the HST data for Styx, 
with the major Styx constraint coming from the LORRI images.
This is not unexpected, as only 55 of the 128 HST images had a useful measurement of Styx,
but it does show the huge value of future astrometric observations of Styx by HST, JWST, and 
other space telescopes.
By comparison, the reduced $\chi^2$ residuals for the HST data versus the PLU055 solution are
4.038 for Nix, 0.521 for Hydra, 1.840 for Kerberos, and 4.272 for Styx.
The solution in Table \ref{tab:elts} thus provides a significant improvement for Nix and Styx 
over PLU055, with the largest improvements for Nix being in the early ACS and WFPC2 data.
This again shows the advantage of longer temporal arcs to improve this solution,
and the need for future astrometric follow up of the Pluto system.

\begin{deluxetable}{c|cccc}
    \tablecaption{
        Mean osculating orbital elements of the small satellites, integrated over 1000 years.
        Columns are: 
        Period of the satellite over the orbital period of Pluto-Charon,
        Semimajor axis of the satellites's orbit relative to the barycenter,
        Eccentricity of the satellite's orbit, and
        Inclination of the satellites's orbit to the plane of the Pluto-Charon orbit.
        \label{tab:elts}
    }
    \tablehead{
        & \colhead{$T/T_{PC}$} & \colhead{$a$ (km)} & \colhead{$e$} & \colhead{$I$ (deg)}
    }
    \startdata
        Styx&3.269409&43171.97&0.024767&0.0411\\
        &$\pm$0.000035&$\pm$0.31&$\pm$0.000008&$\pm$0.0102\\
        Nix&3.994111&49337.76&0.015353&0.0249\\
        &$\pm$0.000003&$\pm$0.02&$\pm$0.000002&$\pm$0.0031\\
        Kerberos&5.127685&58279.64&0.009901&0.4210\\
        &$\pm$0.000013&$\pm$0.10&$\pm$0.000006&$\pm$0.0049\\
        Hydra&6.065662&65186.17&0.008572&0.2819\\
        &$\pm$0.000011&$\pm$0.08&$\pm$0.000006&$\pm$0.0023\\
    \enddata
\end{deluxetable}
The period ratios of Nix and Hydra are 3.994111$\pm$0.000003:1 and 6.065662$\pm$0.000011:1, slightly closer to the 4:1 and 6:1
mean motion resonances than in previously published solutions.
The period ratio for Styx is 3.269409$\pm$0.000035:1 and Kerberos is 5.127685$\pm$0.000013; these are much tighter restrictions 
on the orbits of Styx and Kerberos than have previously been published. 
These values were obtained by sampling the osculating period of the satellites 10,000 times over the course of the 1000 year runs.
All of the orbits have low eccentricities, and Kerberos notably has an eccentricity of 0.009901$\pm$0.000006, well below the
upper limit of e$<$0.02 from \citet{2012ApJ...755...17Y}.
The mean inclinations of Nix and Styx to the Pluto-Charon orbital plane are very small:
0.0249$\pm$0.0031$^\circ$ and 0.041$\pm$0.010$^\circ$, respectively.
Hydra has a larger mean inclination of 0.2818$\pm$0.0023$^\circ$, and Kerberos is even higher at 0.4210$\pm$0.0049$^\circ$.

\section{Discussion}

\begin{figure*}
    \plottwo{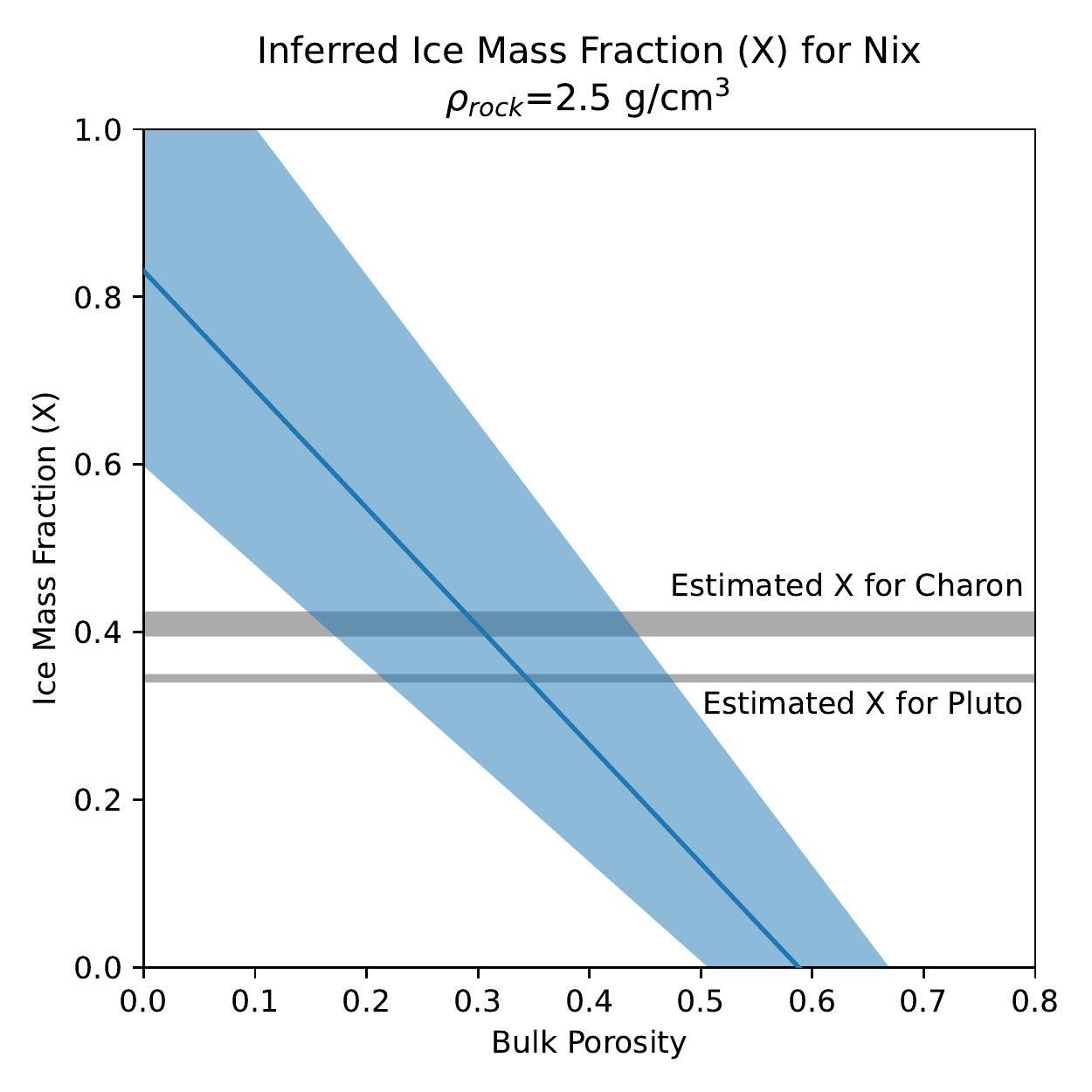}{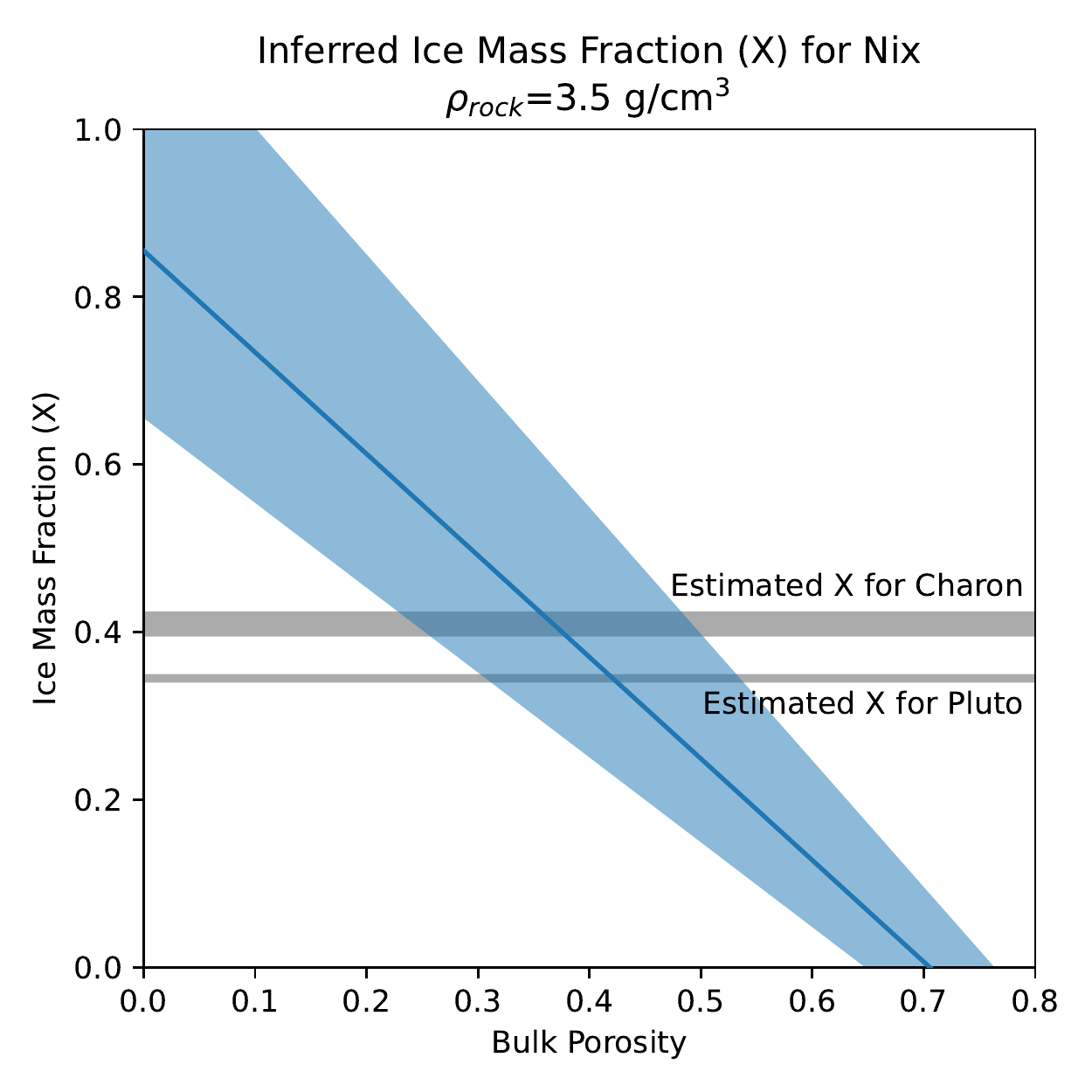}
    \caption{
        Inferred ice fraction for Nix as a function of porosity, 
        using the density estimate in Figure \ref{fig:corner} and Equation \ref{eqn:x}.
        The blue line shows this relationship, with the blue shaded area covering 
        $\pm$1$\sigma$ uncertainty in Nix's density.  
        The grey shaded areas show the estimated ice mass fractions for Pluto and Charon
        from \citet{2017Icar..287....2M}.
        From its high albedo and ice-dominated surface spectra, the ice mass fraction of Nix is not expected to be
        lower than Pluto or Charon, but such values are shown here for completeness.
    }
    \label{fig:nixx}
\end{figure*}

Our results are generally supportive of the concept that the small satellites of Pluto formed out of debris ejected by the giant impact thought to have produced Charon \citep{2006Natur.439..946S,2011AJ....141...35C}.
Our predicted masses for the small satellites are dramatically lower than original estimates by \citet{2008AJ....135..777T}, 
and are also lower than the mean estimates of \citet{2015Icar..246..317B}, particularly for Kerberos, 
and we provide the first usable restriction on the mass of Styx.
\citet{2015AJ....150...11W} found it was very difficult to form the Pluto system with the masses and orbits presented in
\citet{2015Icar..246..317B}.
In particular, \citet{2015AJ....150...11W} found it difficult to form satellites near mean motion resonances.
Our long-term mean period solutions (Table \ref{tab:elts}) have Nix and Hydra closer to resonances, and Kerberos and Styx farther away,
than in the \citet{2015Icar..246..317B} solutions.
This could imply that the small satellites did not initially form near mean motion resonances, but later Nix and Hydra later evolved into
orbits closer to the resonances \citep{2015AJ....150...11W}.

We find that the densities of Nix and Hydra are broadly consistent with $\approx$1.1 g/cm$^3$ 
(see Figure \ref{fig:corner}).
Our results are consistent with the long-term simulations performed by \citet{2022AJ....163..238K} that found that the densities 
of all four small satellites should be $\le$1.4 g/cm$^3$.
The density of Pluto is 1.860$\pm$0.013 g/cm$^3$ and the density of Charon is 1.702$\pm$0.021 g/cm$^3$ \citep{2015Sci...350.1815S}.
Charon is estimated to have $\sim$40\% ice by mass, while
Pluto is estimated to be about one-third ice by mass 
\citep[e.g.,][]{2017Icar..287....2M,2018Icar..309..207B}.
Thus, the small satellites must have a greater ice mass fraction than Pluto 
and Charon, or substantial interior void space (i.e. bulk porosity), or some 
combination of the two.
Porosity must be accounted for, as icy objects of similar sizes 
(e.g. the Saturnian irregular satellites)
are known to be significantly porous \citep{2010Icar..208..395T}.
The relationship between the bulk density $\rho$, 
the mass fraction of ice ($X$), 
the mass fraction of rock (1-$X$), 
and the bulk porosity ($P$, defined as the fraction of the total volume that is void space) 
can be found as:
\begin{eqnarray}
    \rho&=&\frac{M_T}{V_T}=\frac{M_T}{V_{ice}+V_{rock}+V_{void}} ,\\
    \frac{1}{\rho}&=&\frac{M_{ice}}{\rho_{ice}}\frac{1}{M_T}+
    \frac{M_{rock}}{\rho_{rock}}\frac{1}{M_T}+
    {P}{V_T}\frac{1}{M_T} ,\\
    \frac{1}{\rho}&&(1-P)=\frac{X}{\rho_{ice}}+\frac{1-X}{\rho_{rock}} ,\\
    X&=&\left(1-P-\frac{\rho}{\rho_{rock}}\right)
    \left(\frac{\rho}{\rho_{ice}}-\frac{\rho}{\rho_{rock}}\right) \label{eqn:x},
\end{eqnarray}
where $M_T$ and $V_T$ are the total mass and volume, 
$V_{ice}$, $V_{rock}$, and $V_{void}$ are the volumes of the body occupied by ice, rock, and
void space, and $\rho_{rock}$ and $\rho_{ice}$ are assumed densities for the non-porous rock and ice components.

While \citet{2021porterChapter} lists volumes for both Nix and Hydra, the shape model
for Nix is better defined than Hydra, as Nix was imaged along its equator, while Hydra
was only imaged on one hemisphere.
In addition, the shape of Hydra is more complex than Nix, while the shape of Nix is much
closer to a triaxial ellipsoid \citep{2021porterChapter}.
We thus used Nix for the following analysis.  

Figure \ref{fig:nixx} shows Equation \ref{eqn:x} for Nix, assuming the density estimate of 
1.030$\pm$0.204 g/cm$^3$ from Figure \ref{fig:corner},
as well as $\rho_{ice}$=0.92 g/cm$^3$ and $\rho_{rock}$=2.5 or 3.5 g/cm$^3$.
Based on Figure \ref{fig:nixx}, if Nix’s bulk porosity is less than about 30 to 35\%, 
then it is ice-enriched compared to Pluto and Charon, 
while if its bulk porosity is $\sim$30\% to 50\%, 
its composition could be similar to that of Pluto and Charon.
Still higher porosities would imply Nix is more rock-rich than Pluto and Charon, which seems unlikely based on several arguments we describe below.

Most solar system objects similar in size to the small satellites of Pluto 
appear to have high porosities.
Phobos and Deimos are perhaps good analogs, since they are accreted satellites, 
and their porosities are estimated to be in the range 20\% to 60\% 
\citep{2011A&ARv..19...44R,2013LPI....44.1604M}.
The KBO (486958) Arrokoth has been found to have high porosity ($>$50\%) by both
modeling of strength/porosity constraints by \citet{2020Sci...367.6620M}, and 
and by thermal modeling by \citet{2022arXiv220210485U}.
All of the irregularly shaped Saturnian satellites with sizes from 8 to 200 km have densities
between 0.34 and 0.64 g/cm$^3$ \citep{2010Icar..208..395T}, which assuming pure water ice compositions imply
porosities of 36 to 70\%.

Figure \ref{fig:nixx} shows that if the porosity of Nix is greater than about 30\%, it is rock-dominated by mass.   
However, this seems to be inconsistent with the high albedos of the small moons 
\citep{2016Sci...351.0030W}.  
The albedos of Nix and Hydra are $>$50\% \citep{2016Sci...351.0030W}, 
which is quite high compared to the 6-12\% albedo of KBOs in general \citep[and citations therein]{2014ApJ...782..100F}, 
suggesting that the surfaces of Nix and Hydra are predominantly icy.  
\citet{2018Icar..315...30C} found that
the New Horizons LEISA spectra of Nix and Hydra were very similar to Charon,
dominated by water ice, with smaller amounts of a Near IR-dark material 
and Charon-like ammoniated compounds.
\citet{2015Icar..246..360P} found that the majority of impacts onto the small satellites
are erosive, and that ejecta transfer between the small satellites is ineffective
due to both the short dynamical lifetimes of small satellite ejecta ($<$30 Earth years) 
and the small cross-sectional area of the small satellites.
In contrast, the much larger surface gravity of Charon than the small satellites
means that impacts onto it are much more likely to deposit exogenic material 
\citep{2015Icar..258..267G,2019Sci...363..955S}.
Since the majority of those impacts are of low-albedo classical KBOs \citep{2015Icar..258..267G},
the surface of Charon should have become darker over the lifetime of the solar system
from Kuiper Belt impactor pollution,
while the surfaces of the small satellites would not have been contaminated with exogenic dark material.
\textit{However, even erosive impacts that did not deposit their dark material onto the small moons would lead to a darkening of their surfaces over time if material at depth within the moons exposed by the impacts was dark and rocky.}  
Thus, the simplest explanation for how the tiny moons have maintained bright surfaces for billions of years is 
that their interiors are ice-rich like their surfaces 
\citep[e.g.][]{2016Sci...351.0030W,2017Icar..287....2M,2021psnh.book..475C}.

If, as seems most likely, the small satellites have ice-dominated compositions and low bulk porosities,
this may provide an intriguing constraint on their formation, and the formation of the Pluto system.
Formation of an extended debris disk during a Charon-forming giant impact appears most likely when the colliding 
progenitor bodies are partially differentiated, with outer ice shells that overlay mixed ice-rock or hydrated rock
interiors \citep{2011AJ....141...35C}.  
The partially differentiated structure implies outer ice melting during the end stages of progenitor accretion 
\citep{2021psnh.book..475C},
and that ice originating from the progenitor shells would have near solid densities and not be highly porous. 
Ejection of intact fragments of such material into an extended debris disk might then lead to ice-rich but 
relatively low porosity small moons, in contrast to highly porous, ice-rich small KBOs envisioned to result from,
e.g., gentle accretion via the streaming instability 
\citep{2019NatAs...3..808N,2021PSJ.....2...27N},
as seems required for formation of the fragile neck of Arrokoth \citep{2020Sci...367.6620M}.
This suggests that the small satellites of Pluto may be a fundamentally different
type of body than small isolated KBOs, and they hold many secrets of their own.

\section{Summary}

We have produced high-precision orbit and mass estimates for the small satellites of Pluto 
through reanalysis of the HST and New Horizons datasets.
We have shown that these datasets, when combined with a density constraint based on New Horizons resolved 
images and a basic stability constraint, are able to better constrain the masses and densities of Nix and Hydra compared to
previous studies. 
We also provide 1-$\sigma$ upper limits for the masses of Styx and Kerberos; however, the properties of these tiny moons remain poorly constrained.
We find that the densities of Nix and Hydra are close to 1.1 g/cm$^3$,
substantially less than the density of Charon (1.7 g/cm$^3$).
We argue that this, together with the small satellite albedos and surface spectra, is most consistent with the small satellites having a predominantly icy composition
with relatively low porosity, in contrast to the high porosities seen for KBOs and giant planet
satellites of similar sizes.
These results appear broadly consistent with the formation of the small satellites of Pluto from ejecta produced during the giant
impact that formed Charon, with the debris disk from which the small satellites were derived composed primarily of icy
material from the outer shells of the large, partially differentiated progenitors.  
However, challenges with that model remain, notably in explaining the extended orbits of the small moons.

\begin{acknowledgments}
    This work was supported NASA NFDAP grant 80NSSC18K1390.
\end{acknowledgments}

\bibliography{nhmass-arxiv1}
\bibliographystyle{aasjournal}

\end{document}